\documentclass[aps,pre,preprint,superscriptaddress,ruledtabular]{revtex4-2}
\usepackage{appendix}
\usepackage{epsfig}
\usepackage{graphicx}
\usepackage{amsfonts}
\usepackage{amsmath}
\usepackage{mathtools}
\usepackage{psfrag}
\usepackage{amssymb,amsmath,amsthm}
\usepackage{verbatim}
\usepackage{soul}
\usepackage[normalem]{ulem}
\usepackage{color}
\newcommand{\beq}{\begin{equation}}
\newcommand{\eeq}{\end{equation}}
\newcommand{\bea}{\begin{eqnarray}}
\newcommand{\eea}{\end{eqnarray}}
\newcommand{\bwt}{\begin{widetext}}
\newcommand{\ewt}{\end{widetext}}
\allowdisplaybreaks

\begin{document}


\title{Nonlinear acoustic characterization of heterogeneous plasticity in bent aluminium samples}


\author{Carolina Espinoza}
\email[Corresponding author: ]{carolinaaespinozao@uchile.cl} 
\affiliation{Departamento de Sonido, Facultad de Artes, Universidad de Chile, Compa\~n\'ia de Jes\'us 1264, Santiago, Chile.}
\affiliation{Departamento de F\'isica, Facultad de Ciencias F\'isicas y Matem\'aticas, Universidad de Chile, Avenida Blanco Encalada 2008, Santiago, Chile.} 

\author{Vicente Salinas}
\email[]{vicente.salinas@uautonoma.cl} 
\affiliation{Grupo de Investigaci\'on en Física Aplicada, Instituto de Ciencias Qu\'imicas Aplicadas, Facultad de Ingenier\'ia, Universidad Aut\'onoma de Chile, Avenida Pedro de Valdivia 641, Santiago, Chile.}

\author{Makarena Osorio}
\email[]{m.osorio.cornejo@gmail.com} 
\affiliation{Departamento de F\'isica, Facultad de Ciencias F\'isicas y Matem\'aticas, Universidad de Chile, Avenida Blanco Encalada 2008, Santiago, Chile.}

\author{Edgar P\'\i o}
\email[]{edgar.pio@sansano.usm.cl} 
\affiliation{Departamento de Ingenier\'\i a Metal\'urgica y Materiales, Universidad T\'ecnica Federico Santa Maria, Av. Espa\~na 1680, Valpara\'\i so, Chile.}

\author{Claudio Aguilar}
\email[]{claudio.aguilar@usm.cl} 
\affiliation{Departamento de Ingenier\'\i a Metal\'urgica y Materiales, Universidad T\'ecnica Federico Santa Maria, Av. Espa\~na 1680, Valpara\'\i so, Chile.}

\author{Fernando Lund}
\email[]{flund@dfi.uchile.cl} 
\affiliation{Departamento de F\'isica, Facultad de Ciencias F\'isicas y Matem\'aticas, Universidad de Chile, Avenida Blanco Encalada 2008, Santiago, Chile.}

\author{Nicol\'as Mujica}
\email[]{nmujica@dfi.uchile.cl}
\affiliation{Departamento de F\'isica, Facultad de Ciencias F\'isicas y Matem\'aticas, Universidad de Chile, Avenida Blanco Encalada 2008, Santiago, Chile.}



\begin{abstract}
Knowledge of the state of plastic deformation in metallic structures is vital to prevent failure. This is why non-destructive acoustic tests based on the measurement of first order elastic constants have been developed and intensively used. However, plastic deformations, which are usually heterogeneous in space, may be invisible to these methods if the variation of the elastic constants is too small. In recent years, digital image correlation techniques, based on measurements carried out at the surface of a sample, have been successfully used in conjunction with finite element modeling to gain information about plastic deformation in the sample interior. Acoustic waves can penetrate deep into a sample and offer the possibility of probing into the bulk of a plastically deformed material. Previously, we have demonstrated that nonlinear acoustic methods are far more sensitive to changes in dislocation density than linear ones. Here, we show that the nonlinear Second Harmonic Generation method (SHG) is sensitive enough to  detect different zones of von Mises stress as well as effective plastic strain in centimeter-size aluminium pieces. This is achieved by way of ultrasonic measurements on a sample that has undergone a three-point bending test. Because of the relatively low stress and small deformations, the sample undergoes plastic deformation by dislocation proliferation. Thus, we conclude that the nonlinear parameter measured by SHG is also sensitive to dislocation density. Our experimental results agree with numerical results obtained by Finite Element Method (FEM) modeling. We also support the acoustic results by X-Ray Diffraction measurements (XRD).  Although intrusive and less accurate, they also agree with the acoustic measurements and plastic deformations in finite element simulations.
\end{abstract}


\maketitle

\newpage
\section{Introduction}

Plastic strain in metals and alloys usually exhibits a non-uniform distribution. This is true under controlled laboratory conditions due to the anisotropy of the various mechanisms responsible for plastic deformation, such as dislocation slip, twinning, and phase transformation \cite{Sachtleber2002} and, most certainly, for pieces in service due to the actual heterogeneous boundary conditions at play \cite{Wei2022}. Recently, the advent of additive manufacturing has raised awareness of the fact that the fabrication process itself can introduce heterogeneous microstructures \cite{Chen2021,Foehring2021}.

The behavior of aluminium, of particular interest to the automobile and aerospace industries, under plastic strain, has been a subject of special interest. Sachtleber et al. \cite{Sachtleber2002} determined the spatial distribution of plastic surface strains of aluminium polycrystals compressed in a channel die, using an image analysis method to determine the change in surface patterns of the sample under consideration. This work spawned a large body of  subsequent research, particularly using the optical correlations obtained at the surface of a specimen in tandem with finite element modeling to obtain information about the behavior of the innards of a sample for a variety of materials. The same principle can be used substituting light by back-scattered electrons, and falls under  the general label of Digital Image Correlation (DIC) \cite{Demir2021}. In addition to aluminium alloys \cite{Kang2006} examples include $\alpha$-iron \cite{Hoc2003}, Zr and Ti alloys \cite{Heripre2007}, stainless steel \cite{DiGioacchino2007} and Inconel 718 \cite{Hestroffer2022}. The field has been reviewed recently by Weidner and Biermann \cite{Weidner2021}.

While the combination of experimental surface measurements with three dimensional finite element numerical modeling has met with undoubted success, the fact remains that it would be desirable to have an experimental tool capable of penetrating a plastically deformed sample without destroying it, and capable of exploring space-dependent plasticity. In this work we show that such a tool is provided by nonlinear ultrasound. Although the length scales available for exploration compare with wavelength, say in the millimiter to hundreds of microns range, ultrasonic waves penetrate with small, and, in any case controlled, attenuation, in metals and alloys \cite{Hikata:1956}. It is thus reasonable to propose ultrasound as a non-intrusive probe to characterize heterogeneous plastic deformation.

The mechanism we have chosen to obtain heterogeneous plastic deformation is a three-point bending test. This generates sufficient heterogeneity in the sample (as compared for example with a tension test) and is standardized so our results should be easy to replicate. In order to study these heterogeneities, bending tests can provide more information about the expected behavior of a piece in service than, say, uniaxial testing \cite{Westermann2011,Saai2016}. More broadly, bending tests are often used to assess the mechanical performance of aluminium alloy sheets \cite{Sarkar2001,Davidkov2011,Muhammad2019,Oda2022}.

Acoustic methods have been used in crack detection \cite{Muller2005,Payan2007,Payan2014,Haupert2015} and nondestructive evaluation of materials in general \cite{mcskimin1961,chen2007,chanbi2018,Tiwari2018}. In recent years, their application to measure dislocation density in metallic materials has been an active area of research. Advances in theoretical modeling and the development of instrumentation have allowed acoustic measurements to emerge as a quantitative tool for measuring dislocation density in the study of the plastic behavior of metals and alloys \cite{Maurel:2008, Maurel:2005,Mujica:2012,Salinas:2017dg}. In previous work, we have  demonstrated that, in addition, different plastic deformation mechanisms, such as dislocation slip and twinning, can be thereby characterized and identified \cite{Salinas:2022}.  A correlation between changes in nonlinear acoustic parameters and the dislocation density in aluminium and copper samples subjected to different thermomechanical treatments has been observed \cite{Espinoza:2018}. We also demonstrated that the nonlinear acoustic parameter obtained by Second Harmonic Generation (SHG) is more sensitive to changes in dislocation density than the one obtained by Nonlinear Resonance Ultrasound Spectroscopy (NRUS) \cite{Espinoza:2018}, and is thus preferable to probe the plastic behavior of metals and alloys.
 In this work we show that SHG, in combination with finite element modeling, allows us to obtain local measurements of plastic deformation within aluminium specimens subjected to three-point bending tests, obtaining information about the plastic state of the samples at different spatial locations.
 
\section{Experimental and numerical methods}
\subsection{Bending test}
\label{bending}

It is well known that a bending test will induce larger stresses around the pushing points, so it is expected that different levels of plastic deformation will occur in the longitudinal direction of a plastically deformed specimen. In order to demonstrate the sensitivity of SHG in identifying zones with different plastic deformations, we decided to apply a bending test to a metallic sample. The test specimen was fabricated out of commercially 1100 pure aluminium (99.0\% pure) with a geometry defined in the standard ASTM E290-14 for bending test \cite{ASTM2015} with dimensions $16\times32\times200$ mm$^3$. The {  as received}  sample without any mechanical deformation nor thermal treatment was labeled as {Original sample {(OS)}}. Before the bending test, the {OS}  was annealed at $400^{\circ}$C for $125$ hours to remove dislocations introduced during the manufacturing processes. The annealed specimen before the deformation was labeled as {Annealed sample {(AS)}}. Then, a quasi-static three-point bending test was performed using an Instron 3369 machine, with a maximum load capacity of $50$ kN. The applied load was $5$ kN with a velocity of $0.01$ mm/min. During the test, the mandrel was positioned above and at the center of the specimen, and the bottom supports were at $1.0$ cm from each edge. The annealed sample after the deformation was labeled as {Post Bending sample {(PBS)}}. 


\subsection{  Acoustic measurements: speed of shear {and longitudinal} waves}

\begin{figure}[t!]
\begin{centering}
\includegraphics[width = 15 cm]{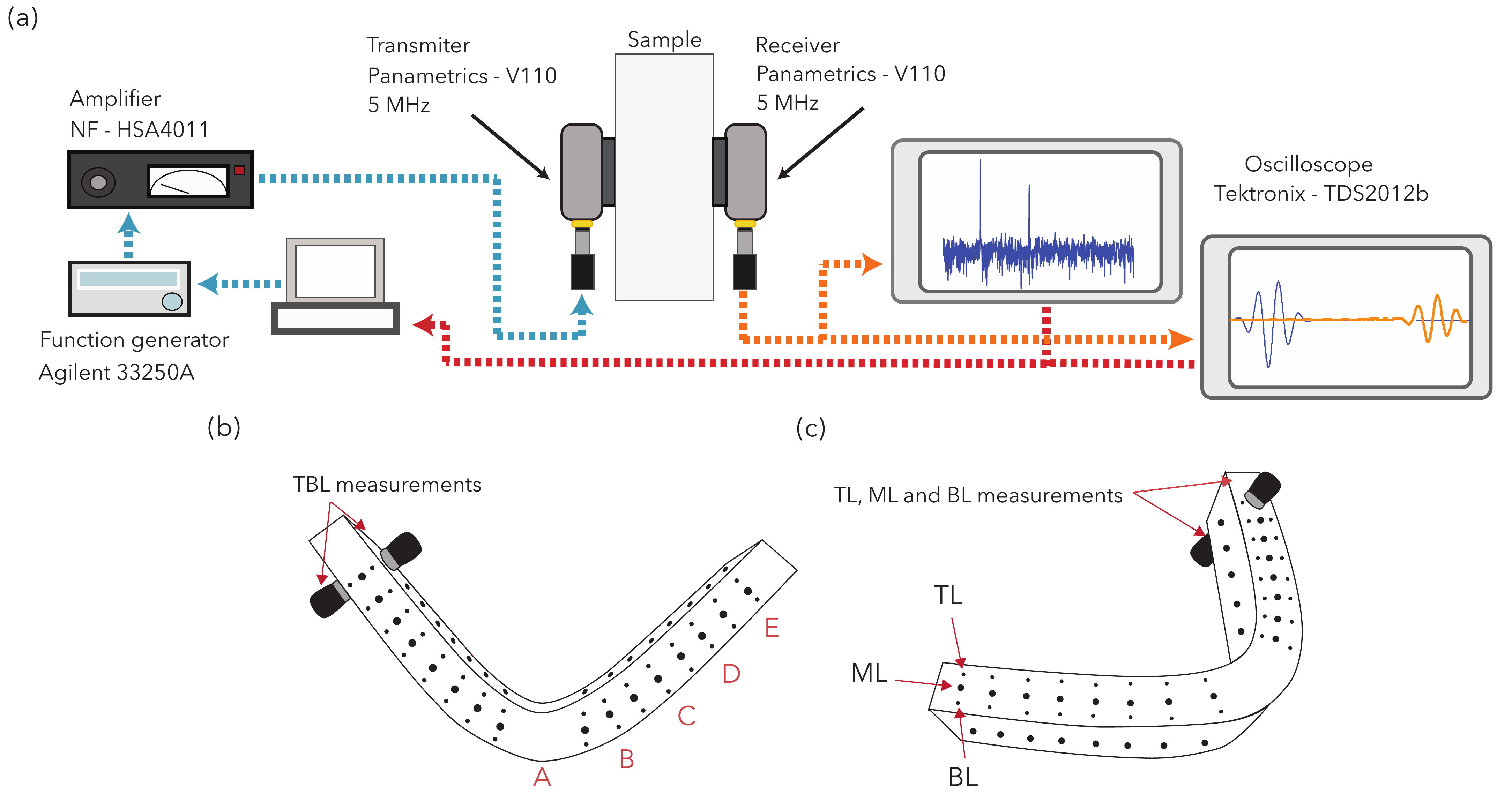}
\par\end{centering}
\caption{Schematic illustration of the experimental setup for SHG measurements. (a) For the TOF measurement, a Gaussian modulated US pulse is generated and amplified. For the SHG measurement, a sinusoidal voltage waveform is generated and amplified. In both cases, the US signal is emitted by a contact transducer, and a second one receives the transmitted signal. For the TOF measurement, both the electrical emission signal and the received US one are acquired by an oscilloscope and these both signals are transferred to a computer. For the SHG measurement, the FFT is computed by an oscilloscope and the first and second harmonic amplitudes are recorded on a computer. Measurements are performed at four different lines: (b) Top-Bottom Line (TBL), (c) Top Line (TL), Middle Line (ML) and Bottom Line (BL). The letters in (b) refer to points where X-ray diffraction measurements were performed.}
\label{setup}
\end{figure}

In this study, the speed of elastic waves were measured with the setup presented in Fig. \ref{setup}(a) using the time of flight (TOF) method: an ultrasonic (US) pulse of carrier frequency $f=3$ MHz that is modulated by a Gaussian function, with a final width of about 5 oscillations, is transmitted into a probe, as in \cite{Salinas:2017dg}. A pair of ultrasonic transducers are used to emit and receive the ultrasonic signal (Panametrics - V110 for longitudinal waves and V156 for transverse waves, both resonant at $5$ MHz, with element diameter $8.8$ mm). The measurements are {performed at points along} the top-bottom line {as} shown in Fig. \ref{setup}(b). The wave speeds are computed by measuring the US pulse time of flight through the sample by cross-correlation between the electrical signal used for US emission and the measured US signal. The distance traveled by the wave, $l$, is obtained at each ultrasonic measurement point using a Mitutoyo micrometer screw, with $0.001$ mm precision.

\subsection{Acoustic measurements: Second Harmonic Generation}
In SHG a second harmonic wave is generated from a propagating monochromatic elastic wave. This is due to the anharmonicity of the elastic material and the presence of micro-structural features such as dislocations. The second harmonic nonlinear response is quantified by the nonlinear parameter \cite{Matlack2014}
\begin{equation}
    \beta=\frac{8}{lk^2}\frac{A_{2\omega}}{A_{\omega}^2},
    \label{eqbeta1}
\end{equation}
where $k$ is the wave number and $l$ is the elastic wave propagation distance. $A_{\omega}$ and $A_{2\omega}$ are the absolute physical displacements of the fundamental and second harmonic waves. The nonlinear parameter $\beta$ has a linear relationship with dislocation density, to a first approximation \cite{Espinoza:2018}. In this study, the SHG method was applied using the setup presented in Fig. \ref{setup}(a): in this case a continuous ultrasonic longitudinal sine wave of frequency $f=3$ MHz is transmitted into a probe. Through Fourier analysis of the received signal, the fundamental ($\bar{A}_{\omega}$) and the second harmonic ($\bar{A}_{2\omega}$) amplitudes, in volts units, are measured. Therefore, using (\ref{eqbeta1}) and $k=2\pi f/v_L$  the nonlinear parameter $\beta'$ is obtained in ${m/V}$ units as 
\begin{equation}
    \beta'=\frac{2v_L^2}{l\pi^2 f^2}\frac{\bar{A}_{2\omega}}{\bar{A}_{\omega}^2},
    \label{eqbeta2}
\end{equation}
where $f$ is the US wave frequency and $v_L$ is the longitudinal speed of wave propagation in aluminium. Finally, {in order} to identify {heterogeneities} in the dislocation density along the samples, four measurement lines were defined:  Top-Bottom Line (TBL, Fig. \ref{setup}(b)), Top Line, Middle Line and Bottom Line (TL, ML and BL, Fig. \ref{setup}(c)). {  The reason for this choice is to explore the different plastic regimes in a three-point bending test: The ``Top'' (concave) region bends in compression, the ``Bottom'' (convex) region bends in tension, and the ``Middle'' region is in between. In this way we explore the fairly complex space dependence of stress that is associated with the plastic bending regime.}

\subsection{X-Ray Diffraction}
In order to support the SHG results, we applied X-Ray Diffraction technique (XRD) to obtain the dislocation density $\Lambda_{XRD}$ at different points of the Post Bending sample{, and compare them with the same measurements of the annealed sample.} For the XRD measurements, we have used the same procedure and equipment reported by Espinoza et al. \cite{Espinoza:2018}. Lattice parameter $a$ and microstrain $\langle\epsilon^2\rangle^{1/2}$, are obtained by Rietveld refinements of the X-ray patterns with the Materials Analysis Using Diffraction (MAUD) software. We use $LaB_6$ ($a=4.1565915(1)$ \AA) as external standard for the determination of instrumental broadening. The calculation of dislocation density $\Lambda_{XRD}$ from the microstructural parameters $a$ and $\langle\epsilon^2\rangle^{1/2}$ is obtained through 
\begin{equation}
\Lambda_{XRD}=\frac{24\pi E}{GF}\frac{\langle\epsilon^2\rangle}{a^2},
\label{eqn_LambdaXRD}
\end{equation}
where $F\approx5$ for FCC materials, $E=74.4\pm1.9$ GPa is the Young's modulus and $G= 28.1\pm0.8$ GPa the shear modulus for Al. These values are calculated as the averages of those reported in \cite{Ogi:2002dg, Lincoln:1967, Vallin:1964}, with error bars given by the computed standard deviations.

\subsection{Finite Element Method Simulations}
{  Finite Element Method (FEM) simulations of {the} bending test have been developed to complement the results obtained experimentally, as well as to obtain parameters that are not possible to measure experimentally.  {The} simulations have been carried out with the finite element technique using the Solid Mechanic module of the COMSOL 5.6 \cite{Comsol2020} program. The simulation reproduces the bending test performed experimentally, with the same probe dimensions (see Sec.\ref{bending}). The simulated material properties are obtained from the COMSOL material library and are complemented by the experimentally measured behavior. \\

The simulation is performed up to the experimentally measured loading limits, which implies that the material, in certain regions, exceeds the yield stress. When this occurs, an isotropic bilinear plasticity model is used, in which Young's modulus, Yield stress and isotopic tangent modulus are required as input variables. {  The bilinear method consists of representing the stress-strain curve as two straight lines intersecting at a point corresponding to creep. The first line, which has zero intercept and ends at creep, corresponds to the elastic zone of the material. The slope of this first line corresponds to its Young's modulus. The second straight line indicates the creep point and ends at the maximum deformation point. The slope of this line, which corresponds to the plastic zone of the material, is called the isotopic tangent modulus.} The values used for these parameters are: Elastic modulus $E=74.0$ GPa 
; Yield stress $\sigma_Y=30.0$ MPa; Isotropic tangent modulus $E_t=2.0$ GPa. These values are obtained from previous tests performed on the same material \cite{Salinas:2017dg}.\\

The boundary conditions of the system are presented in Figure \ref{setupFEM}. On top, at the center of the specimen, a force is exerted on the surface 
\begin{equation}
S\cdot\boldsymbol{n}= \boldsymbol{F_A} = \frac{\boldsymbol{F_{Tot}}}{A}, \nonumber
\end{equation}
where $S$ is the second Piola-Kirchhoff stress tensor, $\boldsymbol{n}$ is the normal vector of the contact surface, $\boldsymbol{F_A}$ is the force per unit area, $\boldsymbol{F_{Tot}}$ is the total applied force and $A$ is the contact surface.  The contact surface used in the simulation corresponds to the contact area between the upper cylinder of the tensile machine and the upper face of the specimen. This area is obtained from the surface deformation experienced by the sample after the experimental test. The latter varies between $0$ and $5$ kN, with $500$ N steps. The simulation is performed quasi-statically, {in agreement with experimental conditions}. At the bottom surface, two stainless steel cylinders are configured as pivots and no displacement is allowed at these points ($\boldsymbol{\bar{u}=0}$, where $\boldsymbol{\bar{u}}$ is the displacement vector in the pivot). The lower pivots are $180$ mm apart.
\begin{figure}[t!]
\begin{centering}
\includegraphics[width = 12 cm]{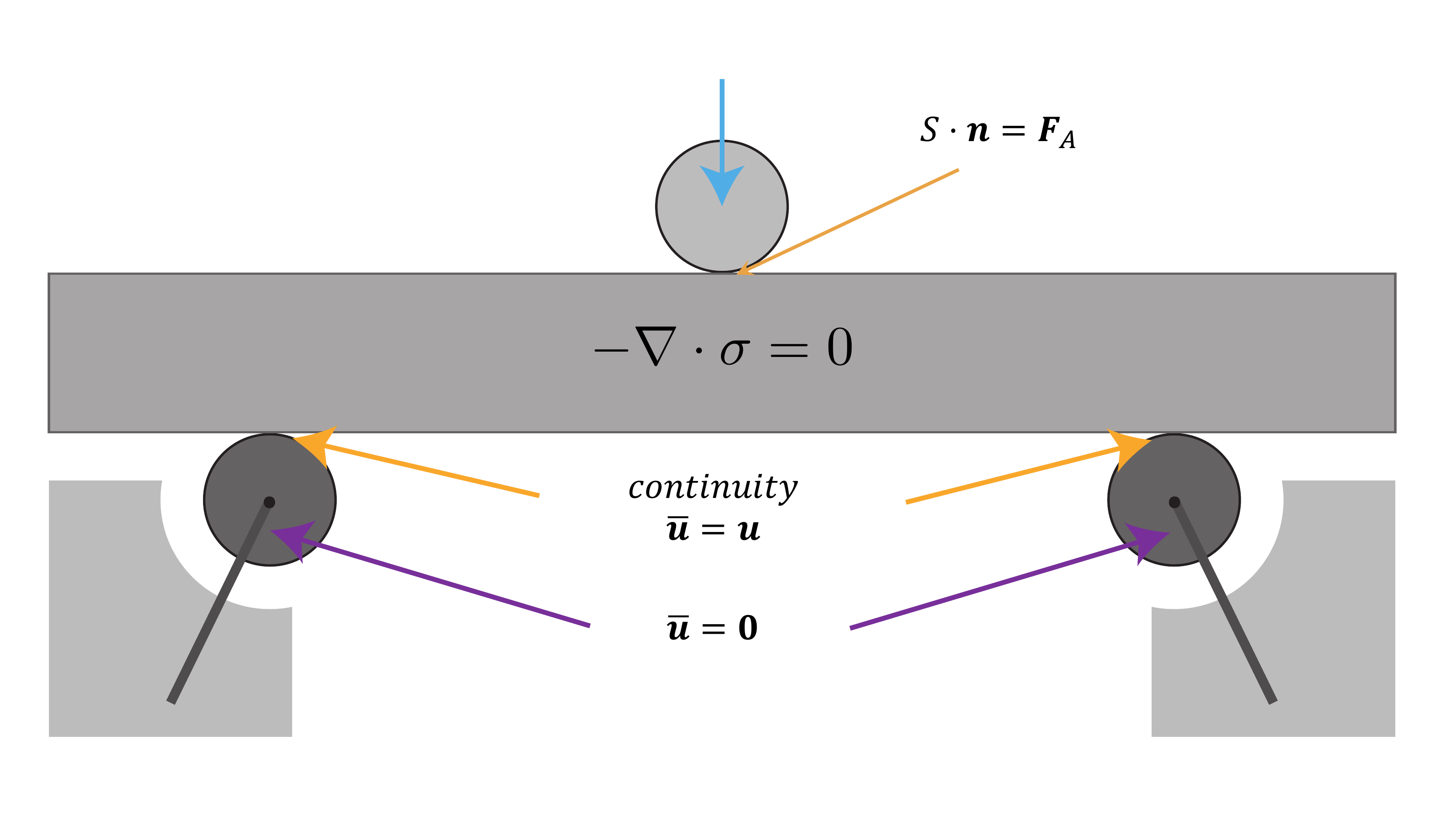}
\par\end{centering}
\caption{Schematic drawing of the boundary conditions used in the 3-point bending test simulation process. The lower pivots have the ``fixed volume constraint'' ($\boldsymbol{\bar{u}=0}$) which does not allow deformation or movement of these pivots. The contact edge condition is continuity, so the deformation of the sample at the contact edge ($\boldsymbol{u}$ is the displacement vector in the sample) with the pivots is zero. The upper cylinder imposes the force on the specimen with the ``boundary load condition'' on the contact surface between the cylinder and the specimen. 
}
 \label{setupFEM}
\end{figure}
From the simulation, the von Mises stress ($\sigma_{VM}$), the plastic strain ($\epsilon_{pe}$) and the strain in the whole specimen are obtained, i.e. as function of position $(x,y,z)$. With this information, and by proper space integration, the averages $\sigma_{VM}(x_i)$ are obtained at the same positions $x_i$ where the nonlinear parameter $\beta'$ is experimentally measured (TBL, TL, ML and BL measurements).}


\section{Results}


\subsection{XRD results}
{  
As in previous work {with polycrystalline aluminium}, the XRD diffraction patterns showed a distribution of crystallite sizes (phases) contributing to each diffraction peak. Each phase has an associated microstrain $\langle \epsilon^2 \rangle^{1/2}$. To obtain the dislocation density $\Lambda_{XRD}$ using Eqn. (\ref{eqn_LambdaXRD}), we use the volume fraction of each phase provided by MAUD as a weight factor. Table \ref{table:XRD} shows the XRD results of the dislocation density at different points of the measurement lines, see Fig. \ref{setup} (b), calculated as a weighted average of the results for different crystallite sizes.} {  This is to be compared with the average dislocation density of the original sample, $\Lambda_{XRD}^{ORI} = 8.12 \pm 2.89 \times 10^7$ mm$^{-2}$, as reported in \cite{Salinas:2017dg} }

\begin{table}[h!]
\caption{XRD measurements of dislocation density $\Lambda_{XRD}$ of the {   Post Bending} sample. Errors for XRD measurements are calculated with the Rietveld refinement results.}
\vspace{1em}
\centering
\begin{ruledtabular}
\begin{tabular}{ c  c  c  c } 
\multicolumn{4}{c}{{   Post Bending} sample}\\
\hline
Measurement point&TL $\frac{\Lambda_{XRD}}{10^8}$ (mm$^{-2}$)& ML $\frac{\Lambda_{XRD}}{10^8}$ (mm$^{-2}$)& BL $\frac{\Lambda_{XRD}}{10^8}$ (mm$^{-2}$)\\
\hline
$A$ &   $1.37\pm0.46$ & $1.95\pm0.84$ & $2.20\pm1.99$ \\

$B$&   $0.57\pm0.24$ & $1.27\pm0.46$ & $0.60\pm0.27$ \\

$C$&   $0.57\pm0.24$ & $1.60\pm0.55$ & $0.59\pm0.69$ \\

$D$&   $0.22\pm0.08$ & $0.64\pm0.18$ & $0.11\pm0.25$ \\

$E$&   $0.43\pm0.10$ & $1.79\pm0.92$ & $0.25\pm0.08$ \\ 
\end{tabular}
\end{ruledtabular}
\label{table:XRD}
\end{table}

\subsection{Simulations}
{  
With the numerical simulation, carried out with the same experimental conditions, the values of deformation and stress at each of the points of the mesh used are obtained. For this simulation the maximum finite element size used is $1$ mm.
Figure \ref{VonMisesSurf} shows an isometric projection view of the specimen in its final condition, i.e. when subjected to a deformation of $50$ mm in the direction of the bending test, for which it was subjected to $5$~kN of force applied in the same direction. Figure \ref{VonMisesSurf} shows that the stress field is heterogeneous in the complete volume, maintaining the symmetry of the bending test. It is observed that along the specimen thickness the stress is not constant, so that the imposed deformation is a function of the position where it is measured.
\begin{figure}[h!]
\begin{centering}
\includegraphics[width = 15 cm]{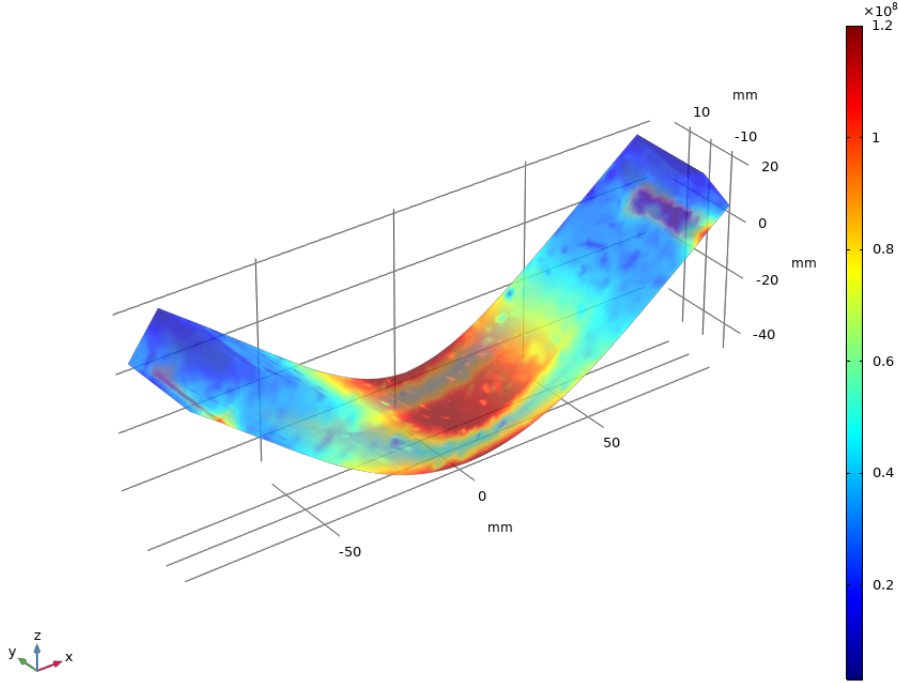}
\par\end{centering}
\caption{Final condition of the specimens subjected to simulated three-point bending test for a force of 5~kN. The color scale shows the Von Misses stress ($\sigma_{VM}$) at each node of the mesh simulation. The points of highest stress are in the contact zone of the lower pivots as well as in the zone where the force is imposed. The spatial distribution of the effective plastic strain ($\varepsilon_{pe}${, not shown}) is qualitatively the same.} 
 \label{VonMisesSurf}
\end{figure}

To obtain the stress to which the specimen is subjected at the {acoustic} measurement points described in Fig~\ref{setup}(c), lines are created in the simulation that cross the specimen starting at the point of contact of the transmitter transducer axis and ending at the point of contact of the receiver transducer. These lines are created with a separation of $2$~mm between them. Along each of these lines, the average von Mises stress is calculated, thus obtaining a quantity that reflects the average deformation along the acoustic wave path in the experimental measurement.\\

The average von Mises stress of each acoustic wave path is presented in Fig~\ref{SimVM}(a). As expected, when plotting this average stress as a function of the x coordinate, it can be seen that its maximum is achieved in the central zone, at the point of force application by the mechanical testing machine, for all the analyzed zones (TL, ML, BL and TBL). Between the coordinates $x=90$ mm and $x=80$ mm (also at $-80$ mm and $-90$ mm) there is a local maximum which correlates with the position of the lower pivots of the bending test. This local maximum is highly attenuated at the TL line, showing that at that height the effect of the pivots is highly reduced. On the ML line, the von Mises stress values are lower than for the other 
lines, but not zero. This is in agreement with what  one would expect of not-too-large deviations from elastic displacements in plane strain,  which indicate that the upper zone of the specimen is subjected to compression, while the lower zone is subjected to tension, and that there is a line, near the center of the specimen, in which the deformation is zero \cite{Dieter:1986}.\\

For the reader's reference, the black segmented lines in Fig~\ref{SimVM}(a) show the position in which the ultrasonic transducers were positioned to obtain the $\beta'$ parameter.\\

{  In the FEM simulation, it is possible to obtain a quantity representing the plastic deformation, i.e. the non-reversible deformation that occurs for the different stresses applied at each finite element node. This quantity is known as "effective plastic strain".  In the process of finite element simulation with the COMSOL structural mechanics module, the 
stress 
is calculated:
\begin{equation}
    \sigma = \sigma_{ext}+C:\varepsilon_{el},
\end{equation}
where $\sigma$ is the stress at each node, $\sigma_{ext}$ is the applied external stress, $C$ is the elastic constants tensor and $\varepsilon_{el}$ the elastic deformation. The total deformation ($\varepsilon$) can be decomposed into the elastic deformation ($\varepsilon_{el}$) and the plastic deformation ($\varepsilon_{pe}$), i.e., $\varepsilon_{el}=\varepsilon - \varepsilon_{pe}$. In this way, a quantity is obtained representing the material's permanent changes, which are associated with the microstructural changes \cite{Dick2018} of the sample subjected to the bending test.
}\\

Fig~\ref{SimVM}(b) shows the average effective plastic strain along each of the acoustic wave paths as a function of the x-coordinate.  It can be seen that for the TL line the effective plastic strain is zero for values of $|x|>70$ mm, thus showing that the pivots do not generate plastic strain at that position. For the ML line, the effective plastic strain values are lower than $0.5\%$, showing that along this line the plastic deformation is small and is close to the zero plastic deformation zone of the bending test. For the TBL and BL lines the behavior is quite similar, presenting local maximum strain in the zone of the lower pivots and a maximum plastic strain in the zone where the force is applied by the testing machine.

\begin{figure}[t!]
\begin{centering}
\includegraphics[width = 7.5 cm]{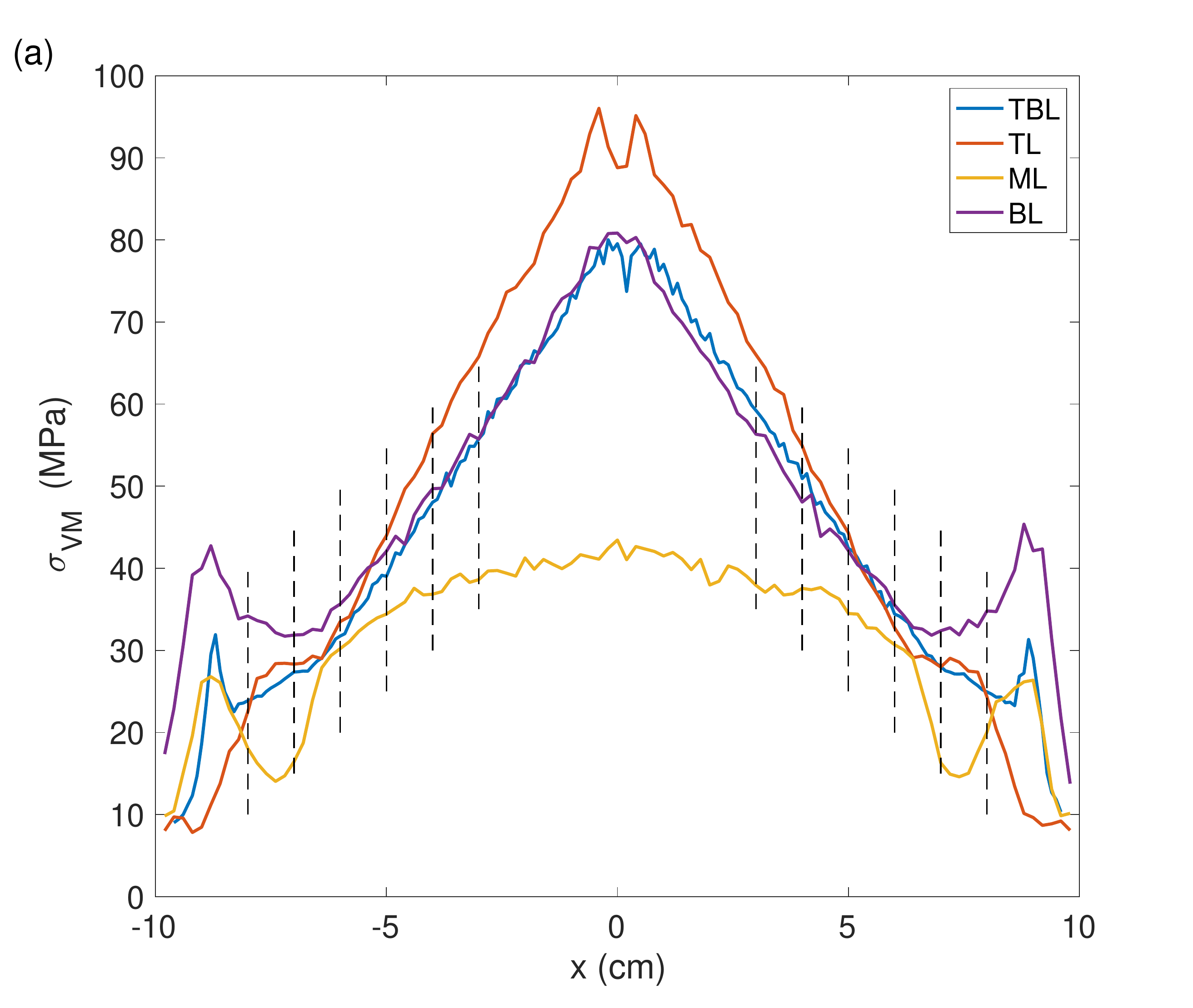}
\includegraphics[width = 7.5 cm]{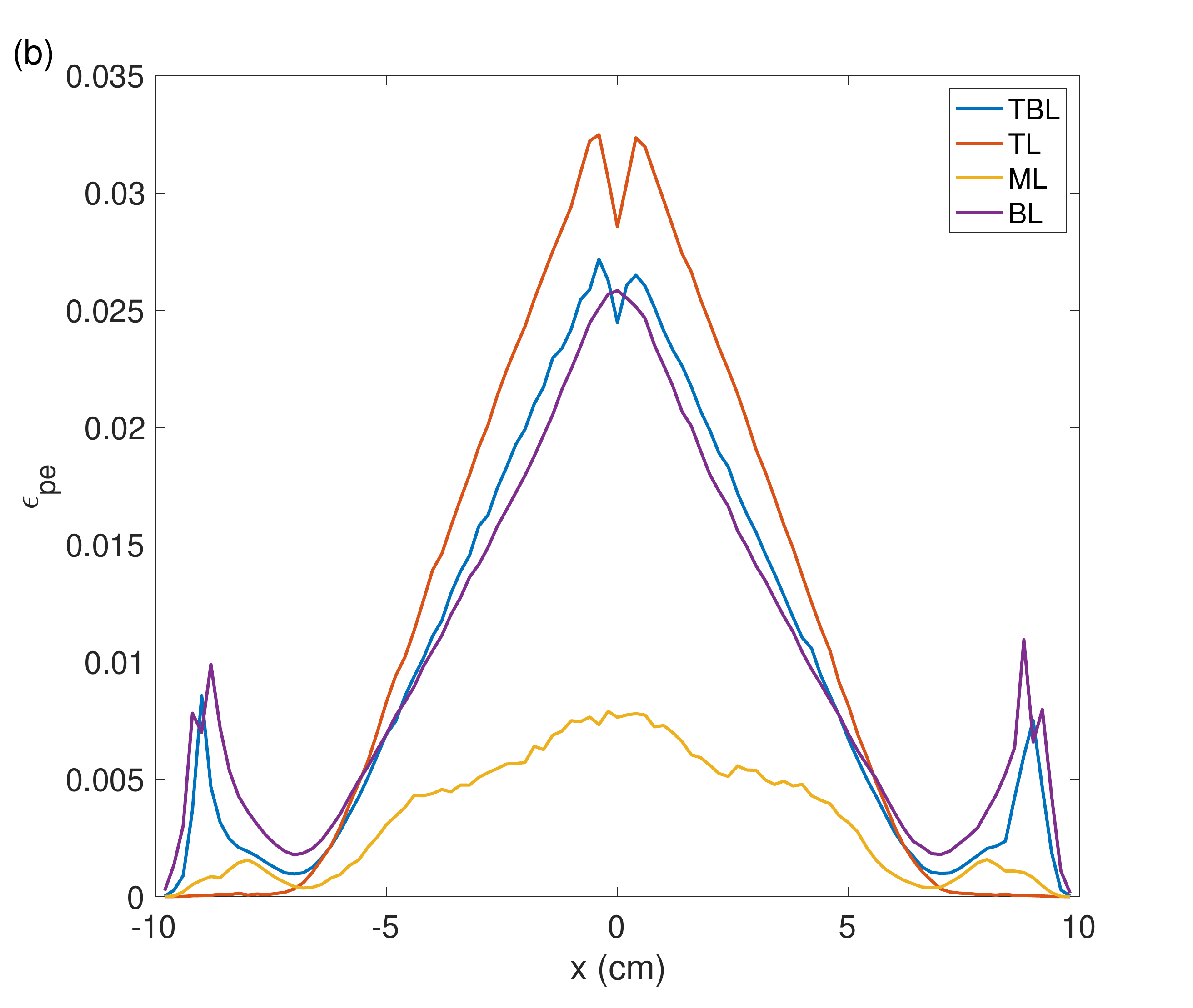}
\par\end{centering}
\caption{Results of simulations of three-point bending test with $5$~kN applied force. (a) Spatial average of von Mises stress along the acoustic wave path as a function of the x-coordinate. The dashed black lines show the locations where the $\beta'$ parameter was experimentally measured. (b) Spatial average of effective plastic strain along the acoustic wave path as a function of the x-coordinate.}
\label{SimVM}
\end{figure}

}

\subsection{Velocity of shear waves for Annealed and Post Bending samples}
\begin{figure}[h]
\begin{centering}
\includegraphics[width = 10 cm]{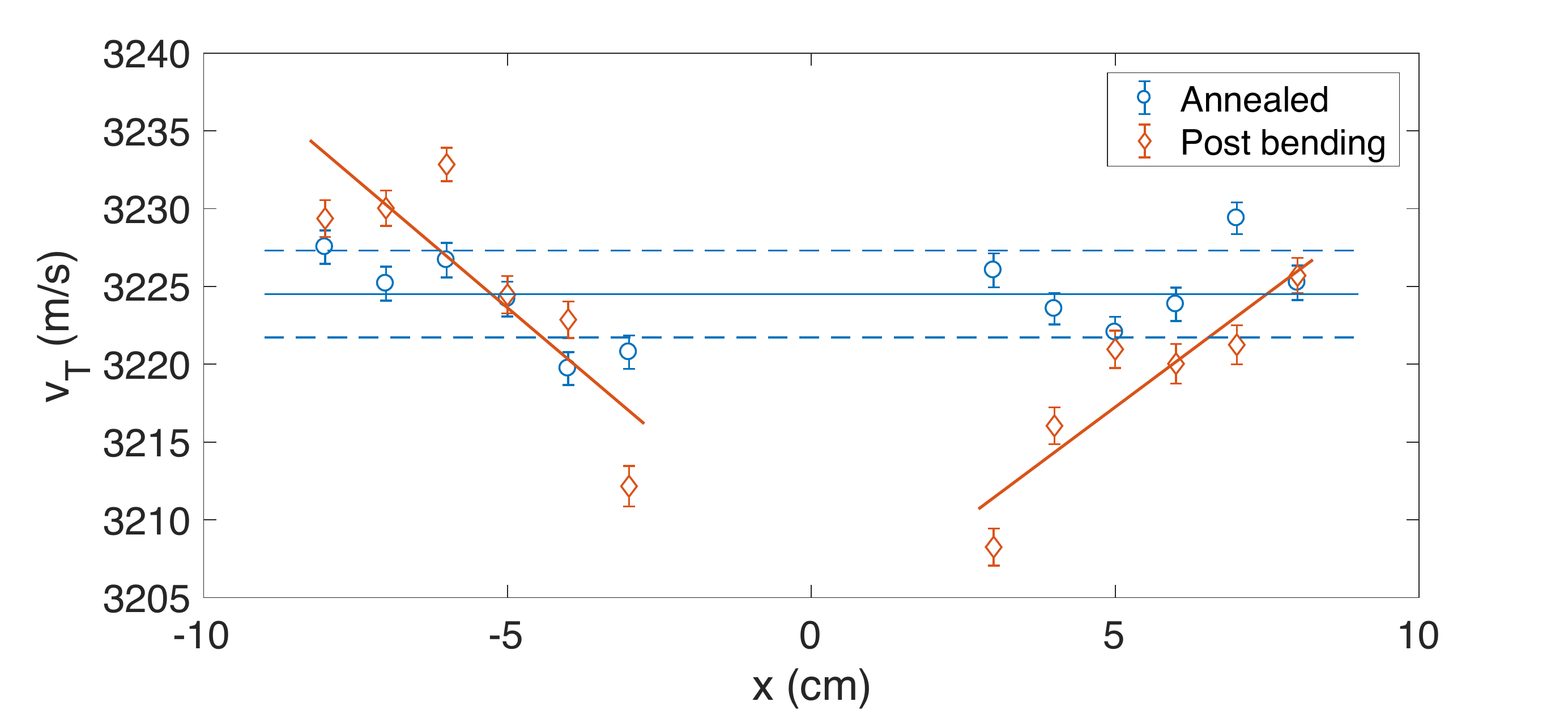}
\par\end{centering}
\caption{{  Shear wave velocity. Blue circles, annealed sample; orange diamonds, post bending sample.} {  The solid blue line represents the shear speed average for the annealed sample, and the dashed blue lines to the standard deviation around this average. Notice that this standard deviation of $2.8$ m/s corresponds to $0.1\%$ of the average value. The red solid lines are guides-to-the-eye showing the linear behavior observed for the shear speed for the post bending sample. In this case, the variation between the maximum and minimum speeds is $\sim 20$ m/s, a difference of about $0.6\%$ respect to the average.} }
\label{vTMaka}
\end{figure}

{  Figure \ref{vTMaka} shows the speed of shear waves measured at different locations of the Original and Post Bending samples, {along} the top-bottom measurement line. Using the formula that relates change in the speed of shear wave  $\Delta v_T$ with change in dislocation density $\Delta \Lambda$ and dislocation length $L$ \cite{Maurel:2005} 
\beq
\frac{\Delta v_T}{v_T}  = -\frac{8\Delta (\Lambda L^2)}{5\pi^4}
\eeq
we obtain, using $L \sim 80$ nm \cite{Salinas:2017dg}, $\Delta \Lambda \sim 0.6 \times 10^8$ mm$^{-2}$. This is similar to the values obtained using XRD, as reported above. The scatter in the data, as well as the experimental uncertainties, preclude a more refined determination of the plasticity behavior of the bent sample as a function of position. As we now show, these shortcomings can be overcome with the measurements of the nonlinear acoustic parameter, in combination with finite element simulations.
}

\subsection{Nonlinear ultrasonics of the original and annealed samples}

\begin{figure}[t!]
\begin{centering}
\includegraphics[width = 15 cm]{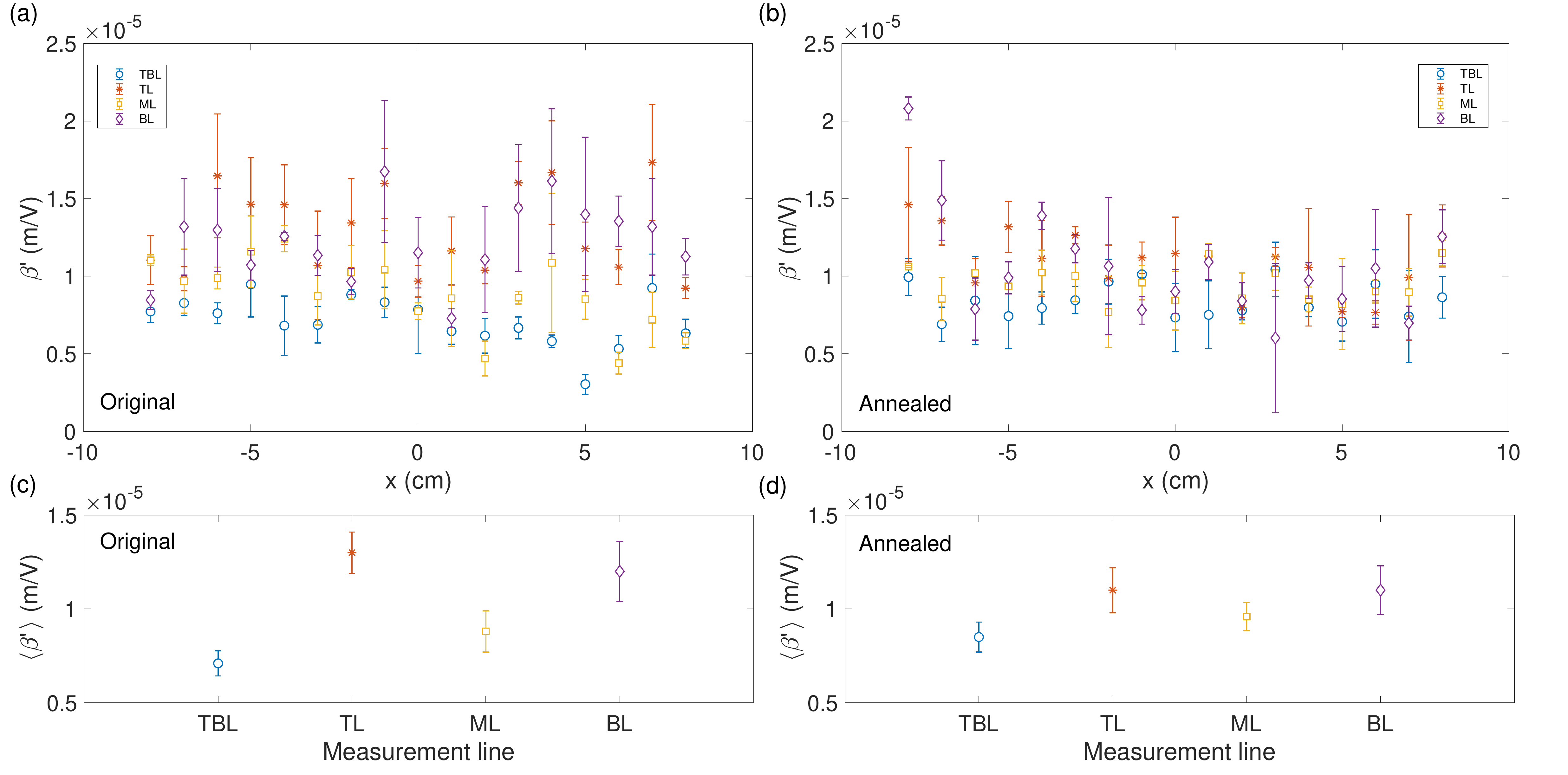}
\par\end{centering}
\caption{Nonlinear acoustic parameter $\beta'$ as function of longitudinal coordinate $x$ for the original (a) and annealed (b) samples for the four measurement lines. Spatial averages for the different lines are shown before and after the annealing procedure in (c) and (d), respectively. A larger dispersion of values is observed for the original sample (a) than for the annealed one (b), {  as expected from the latter's lower dislocation density}. In fact, in each measurement line of the original sample the {  nonlinear} parameter covered clearly separated ranges of values, as shown in (c). The mean value of all the measurements over this sample is $\langle\beta'\rangle_{\rm OS}=1.0\times10^{-5}$  m V$^{-1}$ with a dispersion $\delta\beta'_{\rm OS}=0.3\times10^{-5}$ m V$^{-1}$. {  Panel (d) shows that}  for the annealed sample the nonlinear parameters  tend to collapse to a mean value $\langle\beta'\rangle_{\rm AS}=1.0\times10^{-5}$ m V$^{-1}$ with a dispersion $\delta\beta'_{\rm AS}=0.1\times10^{-5}$ m V$^{-1}$. Although the mean values of the nonlinear parameter $\beta'$ are the same in the original and annealed samples, the dispersion is three times smaller for the annealed sample, which is evidence of the homogenization obtained after thermal treatment.}
\label{original}
\end{figure}

First, using the TOF method we measured an average value of the longitudinal wave speed in the annealed sample,  $\langle v_L\rangle=6071$ m/s, with a dispersion $\delta {v_L}=15$ m/s given by its standard deviation. To calculate the value of $\beta'$ using the expression (\ref{eqbeta2}) we consider this wave speed as a constant before and after the thermomechanical processes. This is justified due to the very small changes it suffers with variations of dislocation density, of the order of $1\%$, much smaller than the variations in $\beta'$ \cite{Espinoza:2018}.

We have characterized $\beta'$ in both the original and annealed sample. The summary of these measurements is presented in Fig. \ref{original}. Before the thermal treatment there is an initial distribution of dislocations that is reflected in an inhomogeneous distribution of $\beta'$. This is clearly demonstrated with the spatial averages shown in panels (c) and (d). In the former, the top and bottom line $x$-coordinate average values are, within experimental errors, the same, and larger than those of the middle line and the top-bottom one. This is most likely due to the manufacturing processes of the probes, where by stress induced machining the surfaces end up with more dislocations, resulting in larger values of $\beta'$ \cite{Espinoza:2018}. On the other hand, in (b) and (d) we show that after the thermal treatment, the nonlinear parameters collapse to a constant value, with a mean value $\langle\beta'\rangle_{\rm AS}=(1.0\pm0.1)\times10^{-5}$ m V$^{-1}$, and a dispersion that is three times smaller compared with the original, pre-annealed sample. This is in good agreement with previous results \cite{Espinoza:2018}. 

\subsection{Post bending ultrasonics and numerical results}
 
\begin{figure}[t!]
\begin{centering}
\includegraphics[width = 10 cm]{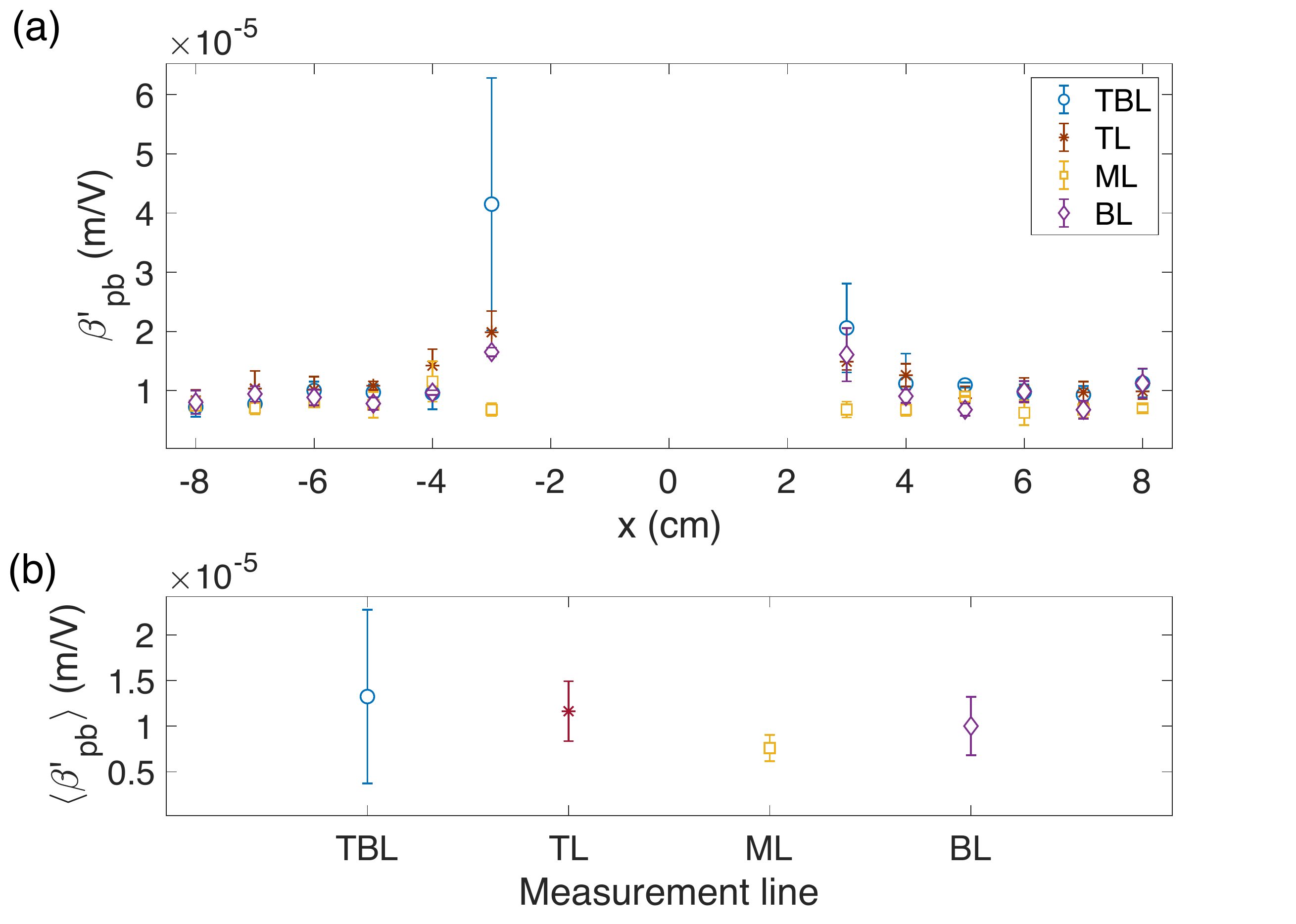}
\par\end{centering}
\caption{(a) Nonlinear acoustic parameter $\beta'$ as function of longitudinal coordinate $x$ of the post bending sample for the four measurement lines. Spatial averages for the different lines are shown in (b). {  Note the difference in the vertical scales compared to Figure \ref{original}}. The parameter $\beta'$ increases for smaller distances to the central bending point, for all lines except the middle one, for which it is almost constant. The mean value of all the measurements over this sample is $\langle\beta'\rangle_{\rm PBS}=1.1\times10^{-5}$  m V$^{-1}$ with a dispersion $\delta\beta'_{\rm PBS}=0.2\times10^{-5}$ m V$^{-1}$, which reflects a small increment with respect to the annealed and original samples. However, the importance here is the spatial sensitivity of $\beta'$, as it clearly increases above the base values in regions of larger stress.}
\label{postbending}
\end{figure}

In order to measure the effect of plastic deformation on the nonlinear acoustic parameter, the SHG method was applied before and after the bending test, on the Annealed and Post Bending samples. In Fig. \ref{postbending} we show the spatial dependence of the nonlinear parameter $\beta'$ as function of the longitudinal coordinate $x$ for the PB sample. For all measurement lines except the middle one (ML), it clearly increases above the base value $1\times 10^{-5}$ m V$^{-1}$ {as it approaches} the central bending point. Concomitantly, it tends to this value as $|x|$ increases. The sample global average, for all lines and all $x_i$ measurement positions, does increase slightly above this value too, as indicated in the figure caption.  

\section{Discussion}

{In order to quantify the changes in the acoustic behavior of the material as a consequence of plastic deformation we consider t}he change in the nonlinear parameter $\beta'$, {which} is calculated as 
\begin{equation}
    \Delta\beta'=\beta'_{\rm PBS}-\beta'_{\rm AS},
    \label{eq:betap}
\end{equation}
where the subfix 'PBS' stands for the Post Bending sample, and 'AS', for the Annealed one. {Since we have $\Delta \beta'$ at various positions of the sample, and, as a result of the FEM simulation, we have both the von Mises stress $\sigma_{VM}$ and the effective plastic strain $\epsilon_{pe}$ at those same locations, it is possible to consider $\Delta \beta'$ as a function of either of the two latter quantities. This we now do.}

\begin{figure}[t!]
\begin{centering}
\includegraphics[width = 7.5 cm]{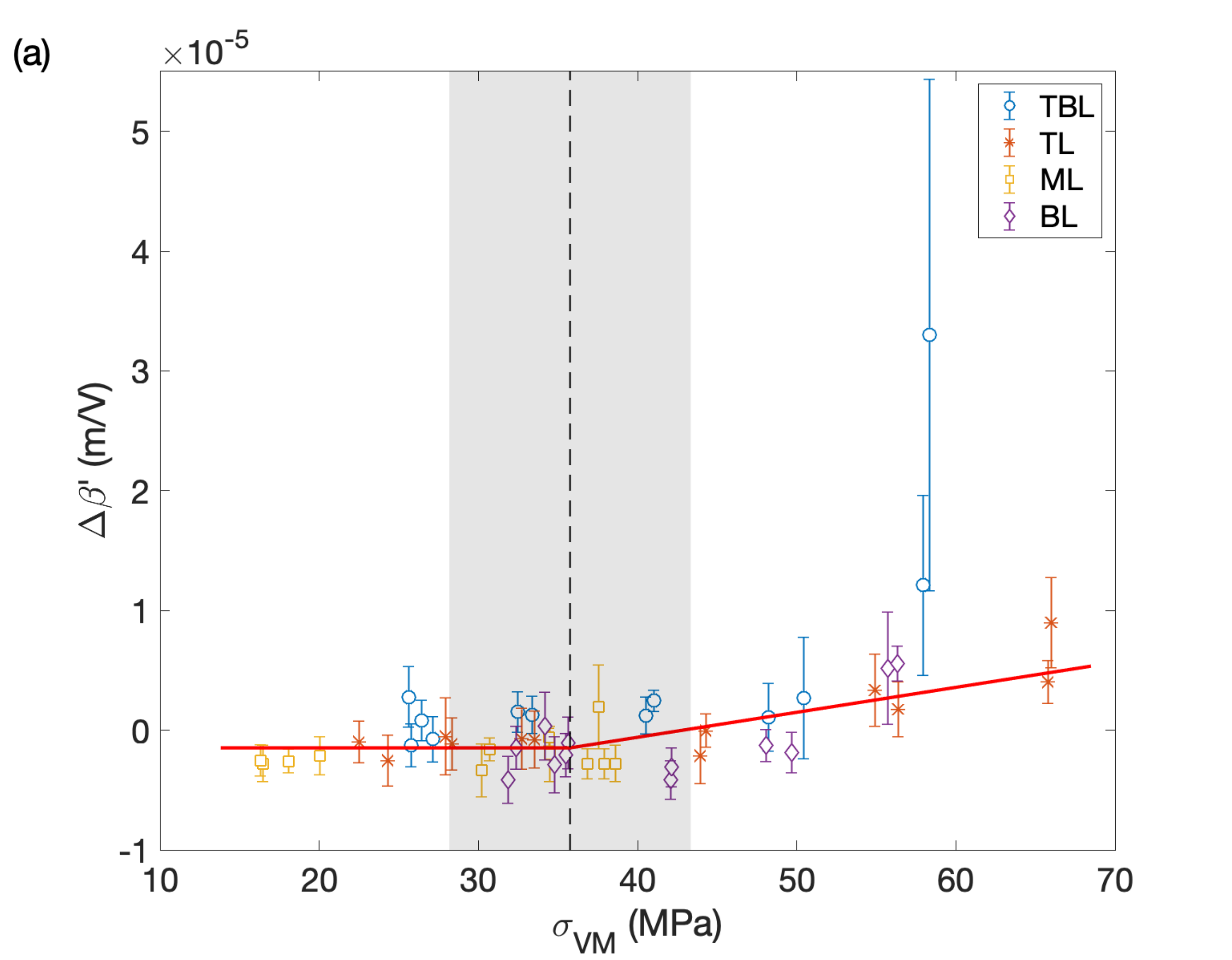}
\includegraphics[width = 7.5 cm]{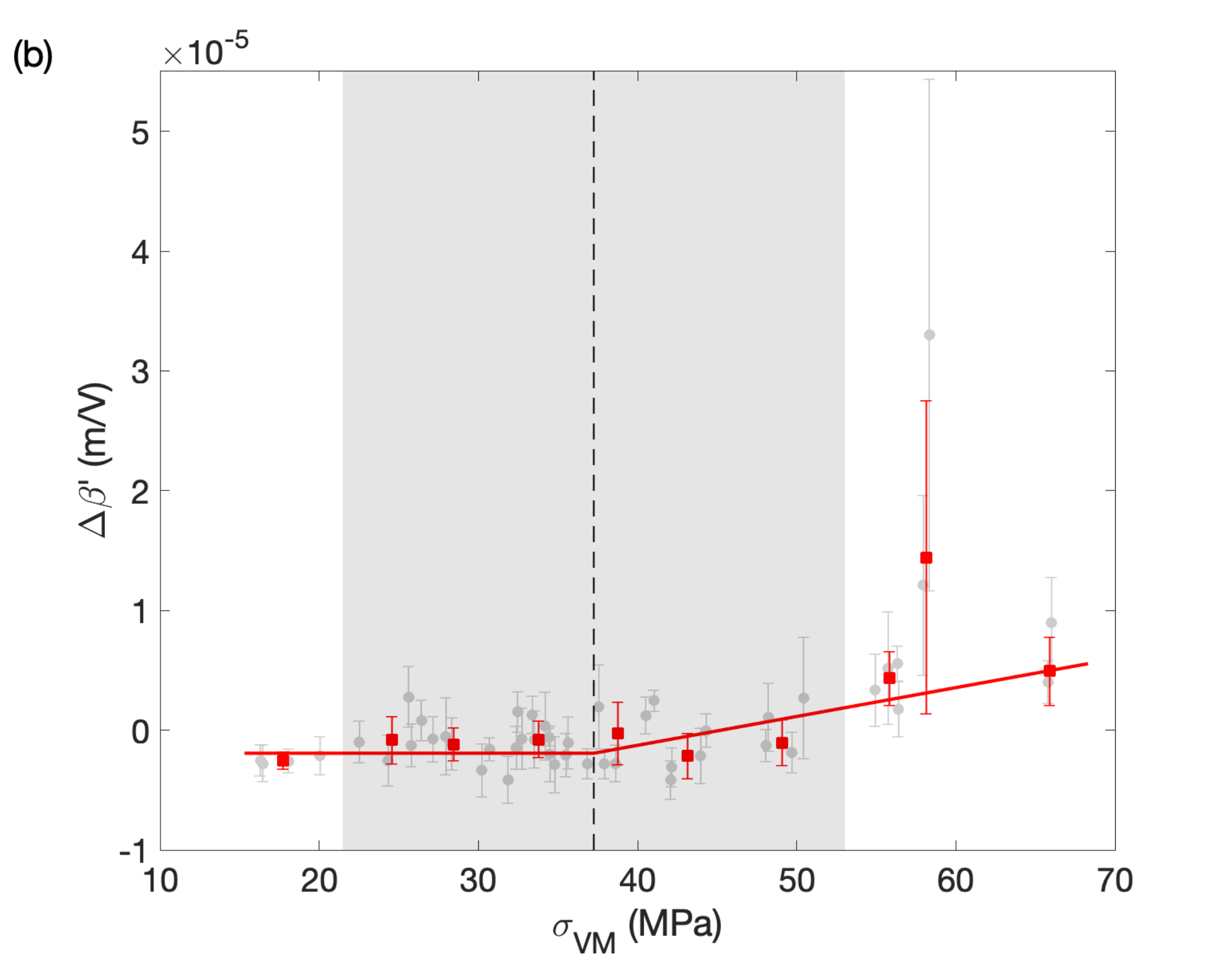}
\par\end{centering}
\caption{Nonlinear parameter difference $\Delta\beta'$ as function of von Mises stress $\sigma_{\rm VM}$ for all measurements. The continuous lines show the linear fits of the form $\Delta\beta' = \Delta\beta'_o$ for $\sigma_{\rm VM}<\sigma_o$ and $\Delta\beta' = \Delta\beta'_o + A(\sigma_{\rm VM}-\sigma_o)$ for $\sigma_{\rm VM}>\sigma_o$. In (a) the fit is done merging data from all measurement lines. In (b), it is done for the window averaged data presented with solid red square symbols. The adjusted values are:  (a) $\Delta\beta'_o=(-1.5\pm0.8)\times10^{-6}$ m V$^{-1}$, $A = (2.1\pm1.2)\times10^{-7}$ m V$^{-1}$MPa$^{-1}$ and $\sigma_o = 35.8.0\pm7.6$~MPa; (b) $\Delta\beta'_o=(-1.9\pm1.2)\times10^{-6}$ m V$^{-1}$, $A = (2.4\pm2.5)\times10^{-7}$ m V$^{-1}$MPa$^{-1}$ and $\sigma_o = 37.3\pm15.8$ MPa. The vertical dashed lines are the fitted threshold stresses $\sigma_o$. The gray shaded regions correspond to the $95\%$ confidence bounds of these fitted parameters.}
 \label{fig:betaVmiss}
\end{figure}

In Fig.~\ref{fig:betaVmiss} we present $\Delta\beta'$ obtained experimentally as function of the von Mises stresses $\sigma_{VM}(x_i)$ obtained numerically, calculated for each measurement line, i.e. averaging{, along the thickness of the sample,} at each measurement position $x_i$. At low stress, the data is consistent with a constant $\Delta\beta'$, very close to zero; a clear deviation is observed above a given stress value, indicated by a vertical dashed line in each panel. This threshold value is more or less consistent with the aluminium's yield stress, although a bit larger than the one of an annealed sample, which is of the order of $30$ MPa. Indeed, for the same material we measured $\sigma_Y \approx 30$, $44$ and $51$ MPa for three consecutive tensile tests on a single probe \cite{Salinas:2017dg}, where the successive increments are due to hardening by dislocation proliferation. In Fig.~\ref{fig:betaVmiss}(a), each measurement line data set is identified with a different symbol and color. {Most of}  the middle line data is below the threshold stress, with $\Delta\beta'\approx 0$. This is expected because this line is between a tensile zone and a compression one, so the local stresses are expected to be very low.  The top and bottom line data are more or less symmetric and deviate clearly from $0$ above a threshold. The top-bottom line data also shows this behavior but with a stronger slope, although errors are much larger for this data set. The continuous red line is the data fit to all measurements merged in one data set, of the form $\Delta\beta' = \Delta\beta'_o$ for $\sigma_{\rm VM}<\sigma_o$ and $\Delta\beta' = \Delta\beta'_o + A(\sigma_{\rm VM}-\sigma_o)$ for $\sigma_{\rm VM}>\sigma_o$. Here, for the fitting procedure we have used weight factors $w = 1/e^2$, where $e$ are the errors for each $\Delta\beta'$. The black dashed vertical line corresponds to the fitted parameter $\sigma_o = 35.8$ MPa.  The gray shaded region corresponds to the $95\%$ confidence bounds of this fitted parameter, which is interpreted as the parameter's error bar; in this case $\pm 7.6$ MPa. The other parameters are given in the figure caption. In Fig.~\ref{fig:betaVmiss}(b), grey solid circles show the merged data set and the solid red squares are window averages: the complete $\sigma_{VM}$ range is divided in 10 bins and all the data that fall into each bin is averaged. In this case, the solid red line is the same functional fit but to this window averaged data set. The fit is also done using weight factors, as defined previously but with the windowed data standard deviation as error. In this case, $\sigma_o = 37.3 \pm 15.8$ MPa, with its error bar  also given by the fitted $95\%$ confidence bounds. It is larger in this case, by a factor 2. 

Here, a natural question is if the von Mises stress is the best quantifier for the local material state. Is it the best correlator to the $\Delta\beta'$ measurements? In other words, if a transition/bifurcation is to be expected/evidenced, is $\sigma_{VM}$ the appropriate diagnose parameter. There are two facts that have to be addressed: (1) The threshold value $\sigma_o$ is about $20\%$ larger than the yield stress of an annealed Al sample; (2) Its associated error bar for the window average procedure is about 2 times the one of the fit using all the measurements merged into one data set. 

\begin{figure}[t!]
\begin{centering}
\includegraphics[width = 7.5 cm]{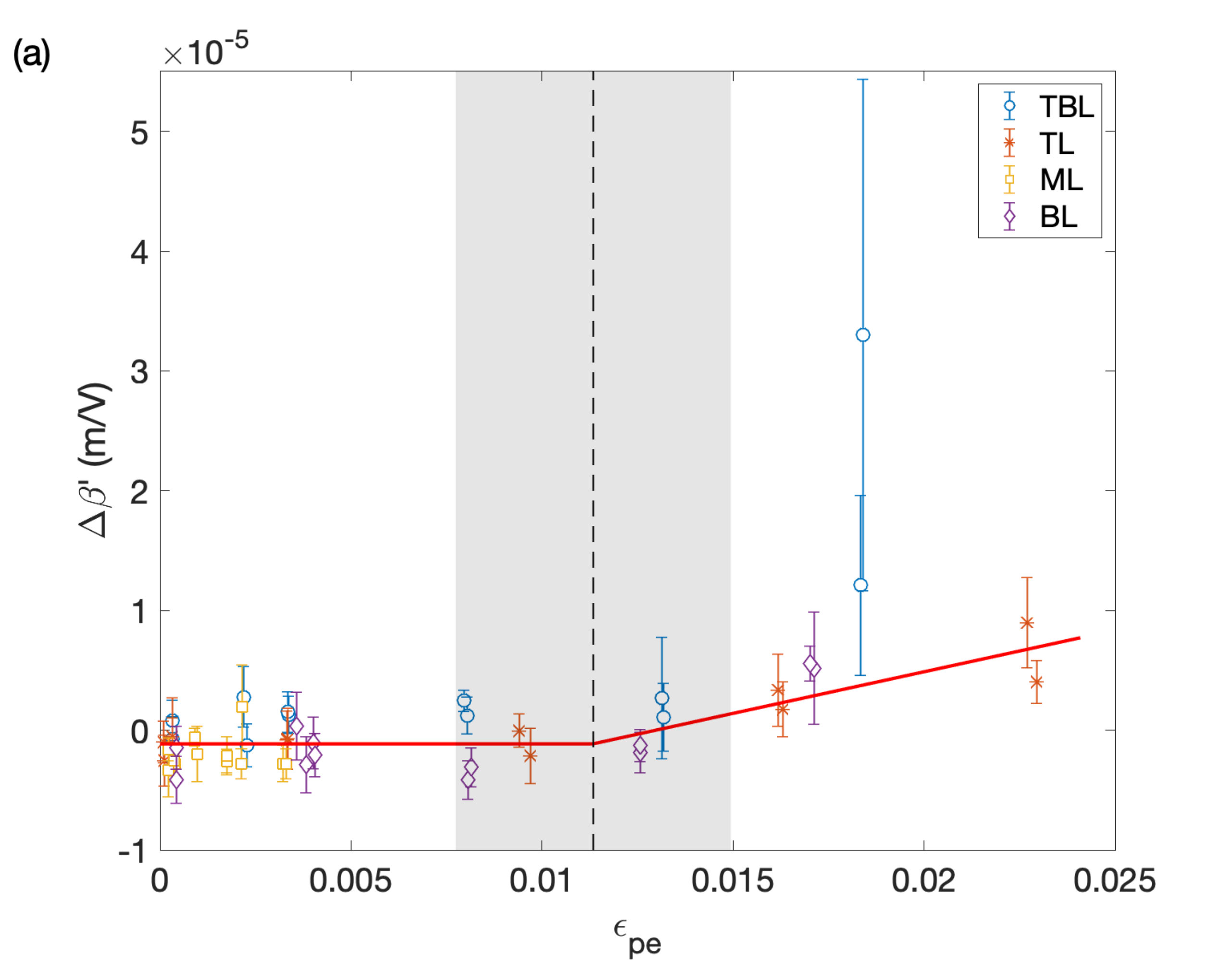}
\includegraphics[width = 7.5 cm]{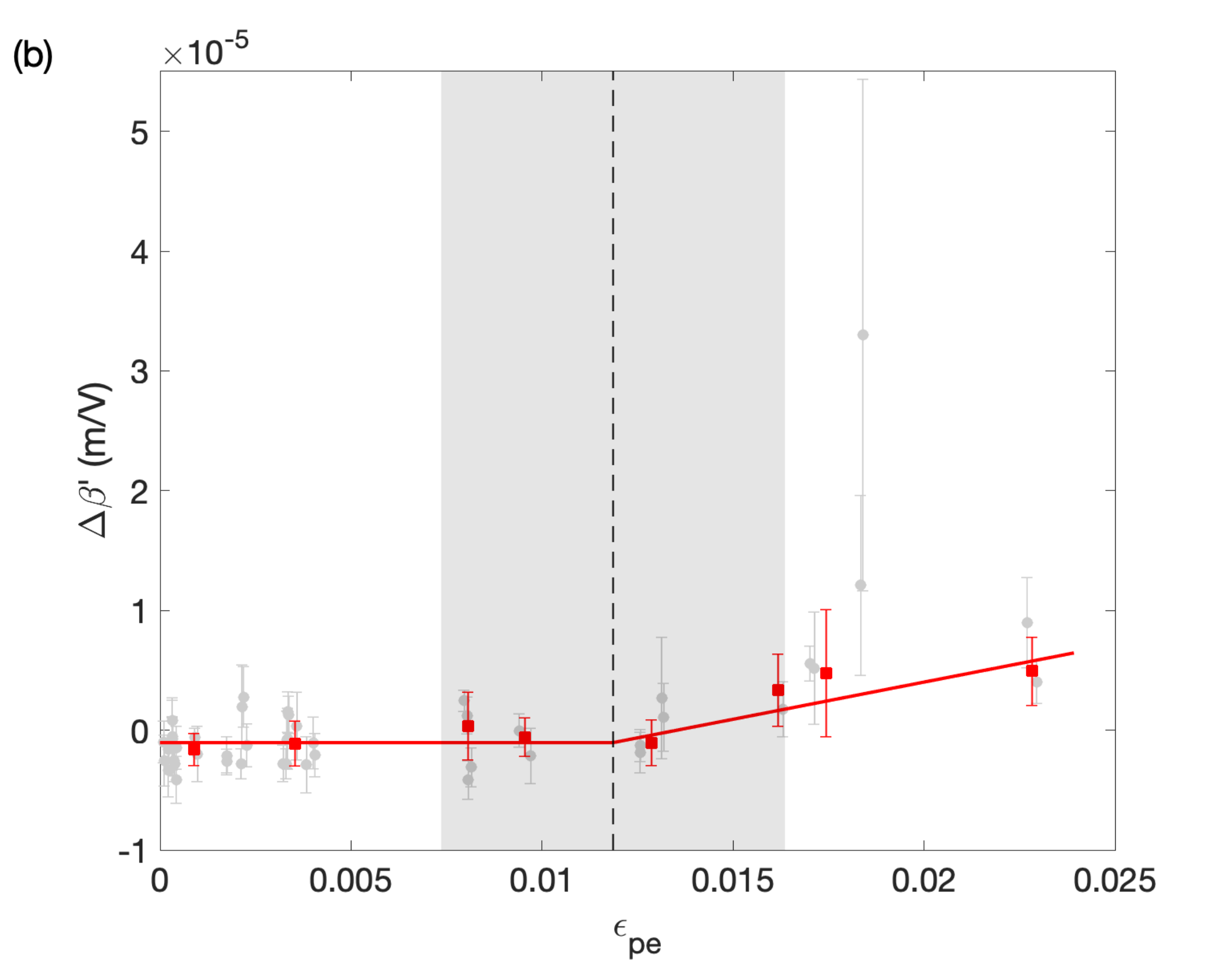}
\par\end{centering}
\caption{Nonlinear parameter difference $\Delta\beta'$ as function of the effective plastic strain $\epsilon_{pe}$. The continuous lines show the linear fits of the form $\Delta\beta' = \Delta\beta'_o$ for $\epsilon_{pe} < \epsilon_o$ and $\Delta\beta' = \Delta\beta'_o + B(\epsilon_{pe} - \epsilon_o)$ for $\epsilon_{pe} > \epsilon_o$. In (a) the fit is done merging data from all measurement lines. In (b), it is done for the window averaged data presented with solid red square symbols. The adjusted values are:\,  (a)
~$\Delta\beta'_o=(-1.1\pm0.6)\times10^{-6}$ m V$^{-1}$, $B = (6.9\pm4.2)\times10^{-4}$ m V$^{-1}$ and $\epsilon_o = (1.13\pm0.36)\times 10^{-2}$; (b) $\Delta\beta'_o=(-1.0\pm1.1)\times10^{-6}$ m V$^{-1}$, $B = (6.2\pm4.3)\times10^{-4}$ m V$^{-1}$ and $\epsilon_o = (1.19\pm0.45)\times 10^{-2}$. The vertical dashed lines are the fitted threshold plastic strains $\epsilon_o$. The gray shaded regions correspond to the $95\%$ confidence bounds of these fitted parameters.}
 \label{fig:betape}
\end{figure}

Thus, we propose as an alternative control parameter the plastic strain $\epsilon_{pe}$. In Fig. \ref{fig:betape} we show $\Delta\beta'$ obtained experimentally as function of $\epsilon_{pe}(x_i)$ obtained numerically, calculated for each measurement line at each ultrasonic measurement point $x_i$. A similar behaviour is observed: a constant $\Delta\beta' = \Delta\beta'_o \approx 0$ is observed for $\epsilon_{pe} < \epsilon_o$ and $\Delta\beta' = \Delta\beta'_o+B (\epsilon_{pe} - \epsilon_o)$ for $\epsilon_{pe} > \epsilon_o$. However, an importance difference exists: for $\epsilon_{pe} < \epsilon_o$, most plastic strain values are concentrated near 0, whereas the von Mises stress values are more evenly distributed between 0 and $\sigma_o$. Here, again all of the middle-line data is concentrated well below the threshold value (at less than one-third), in contrast to the von Mises stress, and the top and bottom line data deviate at a given threshold with a similar slope. Also, again the top-bottom line data deviates with a larger slope but has larger errors. The fitting procedure also uses weight factors given by the errors in each case. The threshold plastic strain $\epsilon_o$ is small, although not $0$, as could be naively expected. Its associated error bars are small, and differ only by a factor of 1.25 between the two procedures in this case. 

\section{Conclusions}
{ 
The results discussed in the previous sections show that the measurement of the acoustic nonlinear parameter $\beta'$ is a reliable measure of plastic deformation within aluminium, and that this measure can be realized with space resolution limited only by the transducers size. First of all, we note that changes in $\beta'$ between annealed and plastically deformed regions are, broadly speaking, of order 50\%. This is a significant change, not difficult to measure. We also note that a universal indicator, long used of plastic behavior is the von Mises stress $\sigma_{\rm VM}$ increasing beyond a given value. Our measurements indicate {that the onset of plastic behavior, as diagnosed by the behavior of $\sigma_{\rm VM}$ is accurately captured by  the behavior of the nonlinear acoustic parameter $\Delta\beta'$. Beyond onset, both quantities are, to a good approximation, linearly related.} {We have also found} that a more reliable experimental indicator {to relate plastic behavior to nonlinear acoustics} is the effective plastic strain $\epsilon_{pe}$. This is not surprising, since $\epsilon_{pe}$ is non-vanishing only within plastically deformed regions. A comparison of the ML values in Figures \ref{fig:betaVmiss}(a) and \ref{fig:betape}(a) illustrates this point. The ML points fall within the region that separates the ``upper'' from the ``lower'' region of the deformed sample where, because of symmetry (which is, admittedly, only approximate in the heavily deformed case) the plastic deformation is smallest. In figure  \ref{fig:betaVmiss}(a) the ML points are broadly distributed for non-vanishing values of $\sigma_{\rm VM}$, whereas in figure \ref{fig:betape}(a) they are clustered near the origin. This is also reflected in the fact that it is possible to ascertain fairly unambiguously that a value of $\Delta \beta' > 0.1$, say, is indicative of $\epsilon_{pe} > 0.01$. A similar relation with $\sigma_{\rm VM}$ is more uncertain. Also, the onset of plasticity as determined by $\epsilon_{pe} > 0.01$ is independent of the yield stress. 

We have also measured the speed of linear shear waves. As noted elsewhere \cite{Espinoza:2018}, this indicator is less sensitive than changes in nonlinear parameters to changes in microstructure and, indeed, our results show changes (see figure \ref{vTMaka}) that are less accurate. Something similar can be said about dislocation measurements using XRD. In addition to its being an intrusive technique, its accuracy (see Table \ref{table:XRD}) is also inferior to nonlinear acoustics. 
}


\section{Acknowledgements}
{  This work was supported by Fondecyt, Chile Grants \#11190900,} {  \#1191179, Fondecyt Postdoctoral Grant \#3200239 and by U. de Chile VID Grant ENL12/22.}

\section{Data Availability}
The raw/processed data required to reproduce these findings cannot be shared at this time as the data also forms part of an ongoing study.




\end{document}